\documentclass[english,notitlepage,superscriptaddress,nofootinbib,twocolumn,prl]{revtex4-2}
\usepackage[T1]{fontenc}
\usepackage[utf8]{inputenc}
\usepackage{amsthm}
\usepackage{amsmath}
\usepackage{amssymb}
\usepackage{graphicx}
\usepackage{braket}
\usepackage[english]{babel}
\usepackage{hyperref}
\hypersetup{colorlinks=true,urlcolor=blue,citecolor = blue}
\usepackage{multirow}
\usepackage{lipsum}

\usepackage{dsfont}
\usepackage{nicefrac}
\usepackage{float}
\usepackage{tikz}
\usepackage{siunitx}
\usepackage{bbold}

\usepackage[newcommands]{ragged2e}
\usepackage{graphicx,caption}
\usepackage{subfigure}
\usepackage[font=small,labelfont=bf,justification=justified]{caption}
\usepackage{xcolor}
\def\be{\begin{equation}}
\def\ee{\end{equation}}

\definecolor{olivegreen}{rgb}{0,0.5,0.1}

\definecolor{darkorange}{rgb}{1, 0.55, 0.0}

\newcounter{secnum}

\newcommand{\prlsection}[2]{%
    \refstepcounter{secnum} 
    \textit{#1.}--\label{#2} 
}

\begin{document}
\title{Witnessing Magic with Bell Inequalities}

\author{R. A. Mac\^{e}do}
\affiliation{Departamento de F\'isica Te\'orica e Experimental, Universidade Federal do Rio Grande do Norte, 59078-970 Natal-RN, Brazil}
\affiliation{International Institute of Physics, Federal University of Rio Grande do Norte, 59078-970, Natal, Brazil}
\author{P. Andriolo}
\affiliation{International Institute of Physics, Federal University of Rio Grande do Norte, 59078-970, Natal, Brazil}
\affiliation{Physics Institute, University of São Paulo, Rua do Matão, 1371, São Paulo, Brazil}
\affiliation{Atominstitut, Technische Universität Wien, Stadionallee 2, 1020 Vienna, Austria}
\author{S. Zamora}
\affiliation{Departamento de F\'isica Te\'orica e Experimental, Universidade Federal do Rio Grande do Norte, 59078-970 Natal-RN, Brazil}
\affiliation{International Institute of Physics, Federal University of Rio Grande do Norte, 59078-970, Natal, Brazil}
\author{D. Poderini}
\affiliation{International Institute of Physics, Federal University of Rio Grande do Norte, 59078-970, Natal, Brazil}
\affiliation{Universit\`a degli Studi di Pavia, Dipartimento di Fisica, QUIT Group, via Bassi 6, 27100 Pavia, Italy}
\author{R. Chaves}
\affiliation{International Institute of Physics, Federal University of Rio Grande do Norte, 59078-970, Natal, Brazil}
\affiliation{School of Science and Technology, Federal University of Rio Grande do Norte, 59078-970 Natal, Brazil}

\begin{abstract}
Non-stabilizerness, or magic, is a fundamental resource for quantum computation, enabling quantum algorithms to surpass classical capabilities. Despite its importance, characterizing magic remains challenging due to the intricate geometry of stabilizer polytopes and the difficulty of simulating non-stabilizer states. In this work, we reveal an unexpected connection between magic and Bell inequalities. Although maximally entangled stabilizer states can violate Bell inequalities and magic is deeply tied to the algebraic structure of observables, we show that tailored Bell inequalities can act as witnesses of magic. This result bridges two key quantum resources, uncovering a novel relationship between the device-independent framework and resource-theoretic properties of quantum computation.
\end{abstract}

\maketitle


\prlsection{Introduction}{sec:intro} Quantum systems exhibit diverse properties with no classical analog, including entanglement \cite{RevModPhys.81.865}, Bell non-locality \cite{RevModPhys.86.419}, and non-stabilizerness \cite{Veitch_2014}, or "magic," each linked to unique operational advantages and physical resources. Among these, non-stabilizerness is particularly relevant for quantum computation, as it underpins the power of quantum algorithms to surpass classical capabilities and allow quantum computers to operate in the fault-tolerant regime \cite{bravyi2005universal,bravyi2016improved, howard2017application}. Stabilizer states, generated by Clifford circuits, even if highly entangled are efficiently simulatable on classical computers \cite{PhysRevA.70.052328}, making them insufficient for universal quantum computation and for achieving quantum advantage. Detecting and quantifying magic is therefore fundamental for understanding the computational potential of quantum systems. However, its characterization remains a significant challenge, stemming from the intricate geometry of stabilizer polytopes \cite{10.1063/5.0222546} and the exponential difficulty of simulating non-stabilizer states on classical platforms.

A similar challenge arises in the characterization of entanglement. Like stabilizer states, separable states also form a convex set \cite{bengtsson2017geometry}, and determining whether a state is separable can be framed as a convex optimization problem often addressed using tools from semi-definite programming (SDP) \cite{PhysRevA.69.022308, PhysRevLett.88.187904}. However, beyond simple cases, SDPs yield only approximations to the separable set \cite{tavakoli2024SDP_review}, and their computational cost escalates rapidly with system size, rendering them impractical for large-scale systems. Another issue, also shared by the characterization of magic, stems from the fact that usual witnesses rely on the precise knowledge of the observable quantities being measured, that is, one has to trust and have a full understanding of the measurement devices. An elegant and general solution to these issues lies in using Bell inequalities \cite{brunner2014bell}. The violation of a Bell inequality provides unambiguous evidence of entanglement within the device-independent (DI) framework \cite{scarani2012device}, eliminating the need for detailed knowledge of the preparation and measurement devices. This raises a compelling question: could Bell inequalities similarly serve as witnesses of magic?

At first glance, it may seem that Bell inequalities are unsuitable for witnessing non-stabilizerness for two key reasons. First, if one considers the paradigmatic CHSH inequality \cite{clauser1969proposed}, one can see that even the maximum violation of it can be achieved by a stabilizer state, thus precluding the typical reasoning used for applying Bell inequalities as a witness of non-classicality. Second, magic is intrinsically tied to the structure of quantum theory, particularly the algebra of observables and the operations that preserve stabilizer states. Non-stabilizerness, as a resource, is defined within a specific mathematical framework, heavily reliant on the commutation relations of observables and the role of the Clifford group. In contrast, Bell nonlocality operates in a device-independent framework, abstracting away details of the measurement apparatus and internal workings of the system, with no explicit reference to uniquely quantum properties like stabilizer structure or Clifford operations. However, as we prove in this paper, tailored Bell inequalities can indeed reveal the presence of magic.

 We begin the paper introducing a general framework for using Bell inequalities as a witness for magic, applicable to arbitrary $d$-dimensional $n$-partite quantum systems. We then demonstrate the application of our approach across various Bell scenarios. Specifically, we employ the tilted CHSH inequality \cite{acin2012randomness} and the CGLMP inequality \cite{PhysRevLett.88.040404} to detect magic in bipartite systems, also showing an example for tripartite systems and discussing its generalization to any number of parties.

\prlsection{Bell inequalities as a resource witness}{sec:bellwitness} Bell's theorem can be seen as proof of the incompatibility between quantum predictions and classical framework of causality \cite{wood2015lesson}. For $n$ spatially separated observers, measuring quantities parameterized by the random variables $X_i$, yielding outcomes $A_i$ with $i=1,\dots,n$, any probability distribution consistent with a classical notion of causality must be expressible as a local hidden variable (LHV) model, represented as  
\begin{equation}
\label{eq:lhv}
p(\mathbf{a}\vert \mathbf{x}) = \sum_{\lambda}p(\lambda) \prod_{i}p(a_i\vert x_i,\lambda),
\end{equation}
with $\mathbf{a}=(a_1,\dots,a_n)$, $\mathbf{x}=(x_1,\dots,x_n)$, with the input and output variable assuming values $a_i=0,\dots,\vert A_i \vert$ and $x_i=0,\dots,\vert X_i \vert$. From this perspective, Bell inequalities represent the constraints imposed by classical causal models \cite{chaves2015unifying}. Their violation reveals the non-classical nature of the underlying physical processes. On the other hand, measurements on $n$-partite separable states, which can be written as $\rho = \sum_{\lambda}p(\lambda)\bigotimes_{i}\rho_{A_i,\lambda} $, are always consistent with a local hidden variable (LHV) model as described in \eqref{eq:lhv}. Thus, the violation of a Bell inequality proves that entanglement is shared between the parties, underscoring the role of these inequalities as a witness to a quantum resource.

More generally, given a set of resource-free states $\mathcal{F}$ we can say that a Bell inequality $I$ will witness that resource if there exists at least one quantum state $\rho$ outside $\mathcal{F}$ that violates $I$. That is, $\delta(\rho)=I(\rho) - I_{\mathcal{F}} > 0$, where 
\begin{equation}
I_{\mathcal{F}} = \max_{\rho \in \mathcal{F},\mathcal{M}} I(\rho, \mathcal{M}),
\end{equation}
with  $I(\rho)=\sum_{\mathbf{a},\mathbf{x}}I^{\mathbf{a}}_{\mathbf{x}}p(\mathbf{a}\vert \mathbf{x})$ describing a $n$-partite Bell inequality and $ p(\mathbf{a}\vert \mathbf{x}) =\mathrm{Tr}\left[\left( M_{a_1\vert x_1} \otimes \dots \otimes M_{a_n\vert x_n}\right)\rho\right]$ being a quantum probability distribution obtained with $\mathcal{M} = \left\{M_{a_i\vert x_i}\right\}$ POVM operators describing measurements over a quantum state $\rho$. 

In this new perspective, we aim to use Bell inequalities as witnesses of non-stabilizerness (magic) \cite{wagner2024certifyingnonstabilizernessquantumprocessors}. Specifically, we seek Bell inequalities $I$ for which $ \delta(\rho) = I(\rho) - I_{\mathcal{F}} > 0 $, where $ \mathcal{F} $ is the set of stabilizer states \cite{gottesman1997stabilizer}. Stabilizer states form a class of quantum states that can be described by a set of stabilizer operators—Pauli operators (or tensor products of Pauli operators)—that leave the state invariant. Alternatively, a stabilizer state can be viewed as the unique $ +1 $ eigenstate of a set of commuting Pauli operators $ \{S_i\} $, such that $ S_i \ket{\psi} = \ket{\psi} $ for all $ i $. The convex hull of these states defines the stabilizer polytope, the full characterization of which remains intractable, even for relatively small systems, due to the exponential growth in the number of stabilizer states, given by $ d^n \prod_{i=1}^{n}(d^{i} + 1) $  for $n$-partite $d$-dimensional systems \cite{Veitch_2014}.

Given a Bell inequality, we thus have to perform the optimization $I_{\mathrm{STAB}} = \max_{\rho \in \mathrm{STAB},\mathcal{M}} I(\rho, \mathcal{M})$ over all states in the stabilizer polytope. For that, we notice that since Bell inequalities are linear, it is enough to consider pure stabilizer states. The optimization can be further simplified if we remember that all stabilizer states are Clifford-locally equivalent to a graph-state $\ket{G}$, that is, $\ket{\psi_{\mathrm{STAB}}}=C_1\otimes \dots C_n\ket{G}$, see \cite{schlingemann2001stabilizer}. Those are stabilizer states defined via a graph $G_=(V,E)$, where $V = \{1,2,\cdots,n\}$ labels the qubits and $E$ are a set of edges. The corresponding group has $n$ generators,
$\mathcal S_G = \langle g_1, g_2, \cdots, g_n\rangle$ with:
\begin{equation}
g_i = X_i \prod_{j \in \mathrm{NN}(i)}Z_j\;,
\end{equation}
where $\mathrm{NN}(i)$ denotes the nearest neighbors of $i$ in $G$. Since the optimization for the optimal value of a Bell inequality is invariant over local unitaries, we see that is enough to optimize over finitely many n-partite graph-states. In the Supplementary Material \cite{supp} we provide details for the generic $d$-dimensional case, but restricting here to the qubit case it is enough to consider projective measurements \cite{device2007acin,masanes2006asymptotic} parametrized as $M_{a_i\vert x_i}=(1/2)(\mathbf{u}_{i,x_i}\cdot\boldsymbol{\sigma}+(-1)^{a_i})$, where $\mathbf{u}_{i,x_i}$ are vectors in the Bloch sphere defining the measurement directions for the $i$-th part and $\boldsymbol{\sigma}=(\sigma_x,\sigma_y,\sigma_z)$. Finding $I_{\mathrm{STAB}}$ is thus equivalent to a polynomial optimization over graph-states, more precisely
\begin{equation}
\label{eq:graphopt}
I_{\mathrm{STAB}}= \max_{G,\mathbf{u}_{i,x_i}} P(\left\{\mathbf{u}_{i,x_i}\right\}),
\end{equation}
where $P(\left\{\mathbf{u}_{i,x_i}\right\})$ is a polynomial defined by the Bell inequality at hand. 

The Clifford-local equivalence has two main advantages. First, it allows an exponential reduction in the number of stabilizer states one has to optimize over, since for $n$ qubits, there are $2^{n\choose 2}$ graph states. For instance, for $n=2$ qubits this allows to reduce from $60$ stabilizer states to $2$ graph states only, while for $n=3$ this reduction is from $1080$ stabilizer states to $8$ graph states. The second feature is the fact that since our witness is optimized over all local unitaries, it allows to the detection of a strong form of non-stabilizerness, since if $I(\rho) > I_{\mathrm{STAB}}$ it follows that $\min_{U_1,\dots,U_n} \mathcal{M}(\bigotimes_i U_i \rho \bigotimes U^{\dagger}_i)>0$ for any measure $\mathcal{M}$ of magic \cite{hamaguchi2024handbook,haug2023scalable}, thus allowing for the verification of non-local magic \cite{cao2024gravitational}.

Having outlined the general framework, we will now apply it to a variety of scenarios.

\prlsection{The tilted CHSH inequality}{sec:TiltedCHSHineq} The simplest Bell scenario involves two parties, each measuring two dichotomic observables. All classical correlations of the form \eqref{eq:lhv} are uniquely characterized (up to relabelings) by the Clauser-Horne-Shimony-Holt (CHSH) inequality \cite{clauser1969proposed}. Unfortunately, the CHSH inequality cannot witness non-stabilizerness. To understand why, consider a pure two-qubit state that, up to local unitaries, can be written as
\begin{equation}
|\psi(\theta)\rangle = \cos(\theta)|00\rangle + \sin(\theta) |11\rangle \quad \theta \in (0, \pi/4)\;.
\label{eq:partent}
\end{equation}
For $\theta = 0$, the corresponding separable state is stabilized by $Z \otimes \mathbb{I}$ and $\mathbb{I} \otimes Z$. Similarly, for $\theta \to \pi/4$, we obtain a maximally entangled state stabilized by $X \otimes X$ and $Z \otimes Z$. However, for all other values of $\theta$, $|\psi(\theta)\rangle$ is a non-stabilizer state. Since the maximum value of the CHSH inequality is attained for a maximally entangled state, this inequality cannot serve as a witness for magic \cite{cepollaro2024harvesting}.

Nicely, as we show next, a minor variation known as the tilted CHSH inequality, originally introduced in the context of randomness certification \cite{acin2012randomness}, also allows for the certification of magic in all pure two-qubit entangled states. The tilted CHSH inequality is given by $I(\alpha) \leq 2+\alpha$ with
\begin{equation}
I(\alpha) \equiv \alpha \langle A^{0}_1\rangle + \sum_{x_1,x_2 \in \{0,1\}}(-1)^{x_1x_2}  \langle A^{x_1}_1 A^{x_2}_2 \rangle\;,
\label{eq:tiltedCHSH}
\end{equation}
with $\langle \cdot \rangle = \mathrm{tr}(\cdot \rho)$, $A^{x_i}_i=M_{0 \vert x_i}-M_{1 \vert x_i}$ and $\langle A^{x_1}_1 A^{x_2}_2 \rangle =\sum_{a_1,a_2} (-1)^{a_1+a_2} p(a_1,a_2 \vert x_1,x_2)$. The parameter $0 \leq \alpha < 2$ is referred to as the \textit{tilting parameter} and when $\alpha=0$, Eq.~(\ref{eq:tiltedCHSH}) simplifies to the standard CHSH inequality.

Since for qubits we can always restrict to projective measurements, the tilted CHSH inequality defines a quadratic polynomial
\begin{align}
I_\mathcal S(\alpha) \equiv&\hspace{0.1cm}  \alpha \mathbf u_{1,0}\cdot \langle \boldsymbol \sigma_1 \rangle \nonumber\\ &+\sum_{x_1,x_2 \in \{0,1\}} (-1)^{x_1x_2} \mathbf u_{1,x_1}^\mathrm T  \langle \boldsymbol \sigma_1 \boldsymbol{\sigma}^\mathrm T_2\rangle\mathbf u_{2,x_2}\;.\label{eq: Stab_Ineq}
\end{align}


Eq.~(\ref{eq: Stab_Ineq}) has as inputs the vectors $\mathbf{u}_{1,0}$, $\mathbf{u}_{1,1}$, $\mathbf{u}_{2,0}$, $\mathbf{u}_{2,1}$ defining the measurement directions and  $\langle \boldsymbol{\sigma}_1\rangle = \langle  G| \boldsymbol{\sigma}_1 | G \rangle$ and $\langle \boldsymbol{\sigma}_1 \boldsymbol{\sigma}^\mathrm T_2 \rangle = \langle  G|\boldsymbol{\sigma}_1 \boldsymbol{\sigma}^\mathrm T_2|G \rangle$, corresponding to a vector and a matrix of Pauli expectation values that depend on graph state $\ket{G}$ under consideration. Therefore, for a fixed graph state, the optimization over the measurement settings can be cast into a quadratically constrained quadratic program (QCQP), which can then be tackled using convex optimization techniques (see Supplemental Material \cite{supp} for details). 
Nicely, however, for the tilted CHSH inequality, we can analytically compute the stabilizer and non-stabilizer bounds.

When Eq.~(\ref{eq: Stab_Ineq}) is evaluated on the Bell state, it yields $2\sqrt 2 $, as the linear part vanishes due to $\langle \boldsymbol \sigma_a \rangle_\mathrm{Bell} =0$ resulting just in the optimization of the CHSH inequality. For the product state, optimization over the measurements gives the local bound, which for this tilted inequality is $2+\alpha$. 

\begin{figure}
    \centering
    \includegraphics[width=1.0\linewidth]{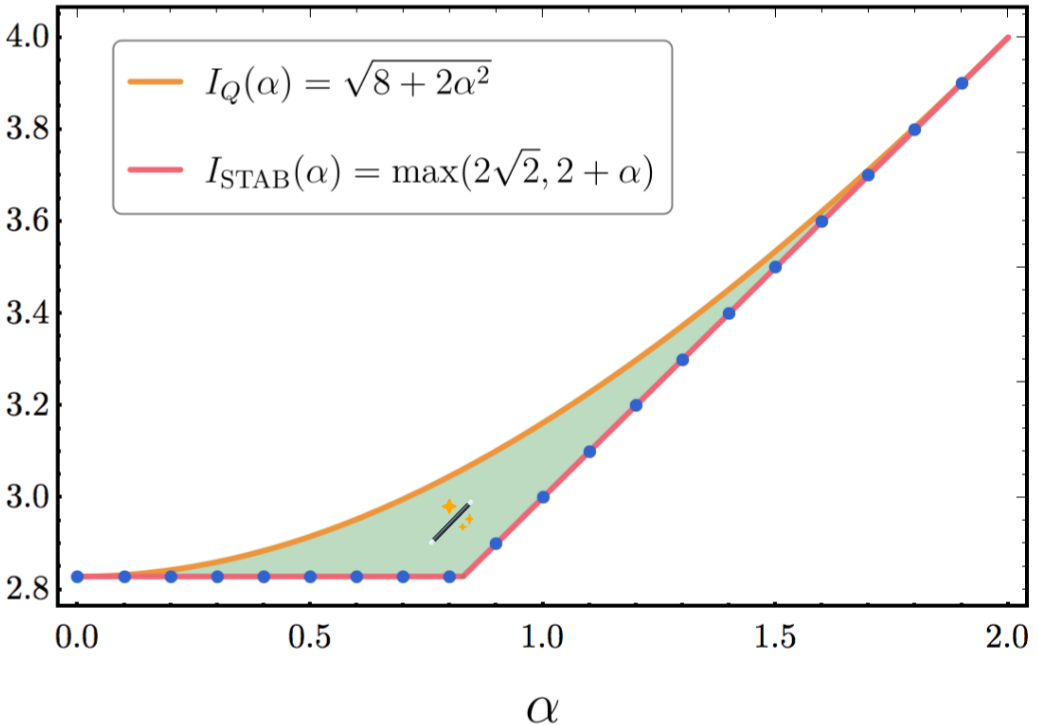}
    \caption{\textbf{Stabilizer/Non-stabilizer gap for the tilted CHSH inequality.} The optimal values of the tilted inequality are shown for all quantum realizations in orange (upper curve) and for stabilizer states in red (lower curve). The blue points correspond to solutions obtained by solving the QCQP optimization. The green region represents the domain where magical states are witnessed.
    }
    \label{fig:STAB22tilted}
\end{figure}

Since for $n=2$ we only have two graph-states, it follows that~\cite{acin2012randomness}
\be
I_\mathrm{STAB}(\alpha) = \max(2\sqrt 2 , 2 +\alpha)\;,
\ee
the first term refeering to a Bell state and the second to a separable state. In turn, allowing for the optimization over nonstabilizer states, we have that $I_Q(\alpha) = \sqrt{8+2\alpha^2}$ \cite{vsupic2020self}, thus demonstrating a gap between the stabilizer and non-stabilizer states for any tilting parameter $\alpha \in [0, 2]$ and any state \eqref{eq:partent} with $\theta \in [0, \pi/4]$ and $\alpha(\theta) = 2(2\tan^2(2\theta) +1)^{-1/2}$. This gap can be seen in Fig. \ref{fig:STAB22tilted}.

\prlsection{Beyond qubits}{sec:GeneralBipartiteScenarios}
Consider a scenario described by $d$-dimensional bipartite quantum states $\rho$ acting on $\mathbb{C}^d \otimes \mathbb{C}^d$. The qudit Pauli operators $X,Z$ are defined by their action on the computational basis $\{\ket{j}\}^{d-1}_{j=0}$, 
\begin{align}
    X\ket{j} = \ket{j+1} , \ Z\ket{j} = \omega^j \ket{j}, \quad \omega\equiv e^{\frac{2\pi i}{d}}.
\end{align}
They are used to construct the generalized Pauli measurements,
\begin{align}
    D_{x,z} = \omega^{2^{-1}xz}X^x Z^z \  ; \ x,z\in \mathbb{Z}_d ,
\end{align}
which forms a basis for the $d \times d$ complex matrices, see \cite{supp} for more details. For dimension $d=2$, one replaces the phase factor by $i$. Define the vector with entries $\left(D_{0,1}, D_{1,0}, \dots , D_{d-1,d-1} \right)$ which has every element of the basis of unitary operators in $\mathbb{C}^d$ except for the identity. This is the generalized version of the single qubit Pauli vector $\boldsymbol{\sigma}$.

Bell inequalities are now written in terms of $d$-dimensional unitary operators $\{A^{x_i}_{i}\}_{i\in\{1,2\}}^{x_{i}\in\{0,1\}}$, satisfying $(A^{x_i}_{i})^{k \dagger} = (A^{x_i}_{i})^{d-k}$, 
and their eigenvalues are roots of unity $\{1, \omega, \omega^2,\dots, \omega^{d-1}\}$. By using a two-dimensional discrete Fourier transform of the joint probabilities $p(a_1,a_2\vert x_1,x_2) = \text{Tr}[(M^{(1)}_{a_1\vert x_1}\otimes M^{(2)}_{a_2\vert x_2})\rho]$,
correlators are given by
\begin{equation}
    \langle (A^{x_1}_1)^k (A^{x_2}_2)^l \rangle = \sum_{a_1,a_2=0}^{d-1}\omega^{a_1 k+a_2 l}p(a_1, a_2\vert x_1, x_2).
\end{equation}
One can then write a generalization of the CHSH inequality for $d$-outcomes, the Collins-Gisin-Lindsen-Massar-Popescu (CGLMP) inequality \cite{PhysRevLett.88.040404}, in terms of the unitary operators as~\cite{salavrakos2017bell}
\begin{align}
    I^{(d)}(A_i^{x_i},\rho) = \sum_{x_1 = x_2}\sum_{l=1}^{d-1} \langle (A_1^{x_1})^l \overline{(A_2^{x_2})^l} \rangle,
\end{align}
where $\overline{(A_2^{x_2})^l} = c_l (A_2^{x_2})^{d-l}  + c_l^* \omega^{(x_2+1) l} (A_2^{x_2+1})^{d-l}$,
 with coefficients $c_l = \sum_{k=0}^{\lfloor\nicefrac{d}{2}\rfloor-1}\left[\alpha_k \omega^{-kl}-\beta_k \omega^{(k+1)l}\right]$. The original form of CGLMP inequality is obtained by choosing $\alpha_k = \beta_k = \nicefrac{(1-2k)}{(d-1)}$.

For the particular case of Hilbert spaces with prime dimensions, the stabilizer states coincide with graph states \cite{Gross_2006,Hostens_2005} thus guaranteeing that the maximal value of the CGLMP operator is achieved by d-dimensional Bell states $\frac{1}{\sqrt{d}}\sum_{i=0}^{d-1}\ket{ii}$. As shown in Table \ref{tabcglmp} since the maximal violation of CGLMP inequality is not achieved by maximally entangled states, this proves that the CGLMP inequality also works as a non-stabilizerness witness.

\begin{table}[H]
\begin{center}
\begin{tabular}{ | >{\centering\arraybackslash}p{2.25cm} || >{\centering\arraybackslash}p{2.5cm} | >{\centering\arraybackslash}p{2.5cm} |  }
 \hline
 \multicolumn{3}{|c|}{Gap between stabilizer and non-stabilizer values} \\
 \hline
  $d$ & $I^{(d)}_\text{STAB}$ & $I^{(d)}_Q$ \\
 \hline
 3 & 2.8729  &  2.9149  \\
 5 & 2.9105 &  3.0157  \\
 7 & 2.9272 &  3.0776  \\
 \hline
\end{tabular}
\caption{\textbf{Stabilizer/Non-stabilizer gap for the CGLMP scenario}. The first column shows the dimension of the stabilizer state under consideration. The second and third columns display the maximum value achievable in the stabilizer ($I^{(d)}_\text{STAB}$ \cite{PhysRevLett.88.040404})  and in the non-stabilizer scenario ($I^{(d)}_Q$ \cite{PhysRevA.65.052325}) when evaluating the corresponding $d$-outcome CGLMP inequality.} 
\label{tabcglmp}
\end{center}
\end{table}

\begin{figure}[h!]
    \centering

\tikzset{every picture/.style={line width=0.75pt}} 

\begin{tikzpicture}[x=0.75pt,y=0.75pt,yscale=-1,xscale=1,scale=0.4]

\draw  [draw opacity=0][fill={rgb, 255:red, 184; green, 233; blue, 134 }  ,fill opacity=0.3 ] (377.13,82.83) .. controls (377.13,44.63) and (408.1,13.67) .. (446.3,13.67) -- (446.3,13.67) .. controls (484.5,13.67) and (515.47,44.63) .. (515.47,82.83) -- (515.47,219.68) .. controls (515.47,257.88) and (484.5,288.85) .. (446.3,288.85) -- (446.3,288.85) .. controls (408.1,288.85) and (377.13,257.88) .. (377.13,219.68) -- cycle ;
\draw  [draw opacity=0][fill={rgb, 255:red, 144; green, 19; blue, 254 }  ,fill opacity=0.3 ] (174.04,244) .. controls (174.04,220.99) and (192.7,202.33) .. (215.71,202.33) -- (249.96,202.33) .. controls (272.97,202.33) and (291.63,220.99) .. (291.63,244) -- (291.63,244) .. controls (291.63,267.01) and (272.97,285.67) .. (249.96,285.67) -- (215.71,285.67) .. controls (192.7,285.67) and (174.04,267.01) .. (174.04,244) -- cycle ;
\draw  [draw opacity=0][fill={rgb, 255:red, 80; green, 227; blue, 194 }  ,fill opacity=0.3 ] (174.67,149.33) .. controls (174.67,126.32) and (193.32,107.67) .. (216.33,107.67) -- (250.58,107.67) .. controls (273.6,107.67) and (292.25,126.32) .. (292.25,149.33) -- (292.25,149.33) .. controls (292.25,172.35) and (273.6,191) .. (250.58,191) -- (216.33,191) .. controls (193.32,191) and (174.67,172.35) .. (174.67,149.33) -- cycle ;
\draw  [draw opacity=0][fill={rgb, 255:red, 184; green, 233; blue, 134 }  ,fill opacity=0.3 ] (173.33,54) .. controls (173.33,30.99) and (191.99,12.33) .. (215,12.33) -- (249.25,12.33) .. controls (272.26,12.33) and (290.92,30.99) .. (290.92,54) -- (290.92,54) .. controls (290.92,77.01) and (272.26,95.67) .. (249.25,95.67) -- (215,95.67) .. controls (191.99,95.67) and (173.33,77.01) .. (173.33,54) -- cycle ;
\draw  [dash pattern={on 4.5pt off 4.5pt}] (3.17,123.13) .. controls (3.17,112.75) and (11.58,104.33) .. (21.97,104.33) -- (84.87,104.33) .. controls (95.25,104.33) and (103.67,112.75) .. (103.67,123.13) -- (103.67,179.53) .. controls (103.67,189.92) and (95.25,198.33) .. (84.87,198.33) -- (21.97,198.33) .. controls (11.58,198.33) and (3.17,189.92) .. (3.17,179.53) -- cycle ;
\draw  [dash pattern={on 4.5pt off 4.5pt}] (362.67,62.27) .. controls (362.67,30.09) and (388.75,4) .. (420.93,4) -- (598.4,4) .. controls (630.58,4) and (656.67,30.09) .. (656.67,62.27) -- (656.67,237.07) .. controls (656.67,269.25) and (630.58,295.33) .. (598.4,295.33) -- (420.93,295.33) .. controls (388.75,295.33) and (362.67,269.25) .. (362.67,237.07) -- cycle ;
\draw  [dash pattern={on 4.5pt off 4.5pt}] (166.67,30.2) .. controls (166.67,15.55) and (178.55,3.67) .. (193.2,3.67) -- (272.8,3.67) .. controls (287.45,3.67) and (299.33,15.55) .. (299.33,30.2) -- (299.33,268.47) .. controls (299.33,283.12) and (287.45,295) .. (272.8,295) -- (193.2,295) .. controls (178.55,295) and (166.67,283.12) .. (166.67,268.47) -- cycle ;
\draw  [draw opacity=0][fill={rgb, 255:red, 74; green, 144; blue, 226 }  ,fill opacity=1 ] (14,170.17) .. controls (14,161.42) and (21.09,154.33) .. (29.83,154.33) .. controls (38.58,154.33) and (45.67,161.42) .. (45.67,170.17) .. controls (45.67,178.91) and (38.58,186) .. (29.83,186) .. controls (21.09,186) and (14,178.91) .. (14,170.17) -- cycle ;
\draw  [draw opacity=0][fill={rgb, 255:red, 74; green, 144; blue, 226 }  ,fill opacity=1 ] (62.67,169.5) .. controls (62.67,160.76) and (69.76,153.67) .. (78.5,153.67) .. controls (87.24,153.67) and (94.33,160.76) .. (94.33,169.5) .. controls (94.33,178.24) and (87.24,185.33) .. (78.5,185.33) .. controls (69.76,185.33) and (62.67,178.24) .. (62.67,169.5) -- cycle ;
\draw  [draw opacity=0][fill={rgb, 255:red, 74; green, 144; blue, 226 }  ,fill opacity=1 ] (38.33,130.17) .. controls (38.33,121.42) and (45.42,114.33) .. (54.17,114.33) .. controls (62.91,114.33) and (70,121.42) .. (70,130.17) .. controls (70,138.91) and (62.91,146) .. (54.17,146) .. controls (45.42,146) and (38.33,138.91) .. (38.33,130.17) -- cycle ;
\draw  [draw opacity=0][fill={rgb, 255:red, 74; green, 144; blue, 226 }  ,fill opacity=1 ] (406.13,166.9) .. controls (406.13,158.16) and (413.22,151.07) .. (421.97,151.07) .. controls (430.71,151.07) and (437.8,158.16) .. (437.8,166.9) .. controls (437.8,175.64) and (430.71,182.73) .. (421.97,182.73) .. controls (413.22,182.73) and (406.13,175.64) .. (406.13,166.9) -- cycle ;
\draw  [draw opacity=0][fill={rgb, 255:red, 74; green, 144; blue, 226 }  ,fill opacity=1 ] (455.13,166.57) .. controls (455.13,157.82) and (462.22,150.73) .. (470.97,150.73) .. controls (479.71,150.73) and (486.8,157.82) .. (486.8,166.57) .. controls (486.8,175.31) and (479.71,182.4) .. (470.97,182.4) .. controls (462.22,182.4) and (455.13,175.31) .. (455.13,166.57) -- cycle ;
\draw  [draw opacity=0][fill={rgb, 255:red, 74; green, 144; blue, 226 }  ,fill opacity=1 ] (430.47,126.9) .. controls (430.47,118.16) and (437.56,111.07) .. (446.3,111.07) .. controls (455.04,111.07) and (462.13,118.16) .. (462.13,126.9) .. controls (462.13,135.64) and (455.04,142.73) .. (446.3,142.73) .. controls (437.56,142.73) and (430.47,135.64) .. (430.47,126.9) -- cycle ;
\draw  [draw opacity=0][fill={rgb, 255:red, 74; green, 144; blue, 226 }  ,fill opacity=1 ] (192.67,264.17) .. controls (192.67,255.42) and (199.76,248.33) .. (208.5,248.33) .. controls (217.24,248.33) and (224.33,255.42) .. (224.33,264.17) .. controls (224.33,272.91) and (217.24,280) .. (208.5,280) .. controls (199.76,280) and (192.67,272.91) .. (192.67,264.17) -- cycle ;
\draw  [draw opacity=0][fill={rgb, 255:red, 74; green, 144; blue, 226 }  ,fill opacity=1 ] (241.67,263.83) .. controls (241.67,255.09) and (248.76,248) .. (257.5,248) .. controls (266.24,248) and (273.33,255.09) .. (273.33,263.83) .. controls (273.33,272.58) and (266.24,279.67) .. (257.5,279.67) .. controls (248.76,279.67) and (241.67,272.58) .. (241.67,263.83) -- cycle ;
\draw  [draw opacity=0][fill={rgb, 255:red, 74; green, 144; blue, 226 }  ,fill opacity=1 ] (217,224.17) .. controls (217,215.42) and (224.09,208.33) .. (232.83,208.33) .. controls (241.58,208.33) and (248.67,215.42) .. (248.67,224.17) .. controls (248.67,232.91) and (241.58,240) .. (232.83,240) .. controls (224.09,240) and (217,232.91) .. (217,224.17) -- cycle ;
\draw  [draw opacity=0][fill={rgb, 255:red, 74; green, 144; blue, 226 }  ,fill opacity=1 ] (192.67,168.5) .. controls (192.67,159.76) and (199.76,152.67) .. (208.5,152.67) .. controls (217.24,152.67) and (224.33,159.76) .. (224.33,168.5) .. controls (224.33,177.24) and (217.24,184.33) .. (208.5,184.33) .. controls (199.76,184.33) and (192.67,177.24) .. (192.67,168.5) -- cycle ;
\draw  [draw opacity=0][fill={rgb, 255:red, 74; green, 144; blue, 226 }  ,fill opacity=1 ] (241.67,168.17) .. controls (241.67,159.42) and (248.76,152.33) .. (257.5,152.33) .. controls (266.24,152.33) and (273.33,159.42) .. (273.33,168.17) .. controls (273.33,176.91) and (266.24,184) .. (257.5,184) .. controls (248.76,184) and (241.67,176.91) .. (241.67,168.17) -- cycle ;
\draw  [draw opacity=0][fill={rgb, 255:red, 74; green, 144; blue, 226 }  ,fill opacity=1 ] (217,128.5) .. controls (217,119.76) and (224.09,112.67) .. (232.83,112.67) .. controls (241.58,112.67) and (248.67,119.76) .. (248.67,128.5) .. controls (248.67,137.24) and (241.58,144.33) .. (232.83,144.33) .. controls (224.09,144.33) and (217,137.24) .. (217,128.5) -- cycle ;
\draw  [draw opacity=0][fill={rgb, 255:red, 74; green, 144; blue, 226 }  ,fill opacity=1 ] (191.67,72.83) .. controls (191.67,64.09) and (198.76,57) .. (207.5,57) .. controls (216.24,57) and (223.33,64.09) .. (223.33,72.83) .. controls (223.33,81.58) and (216.24,88.67) .. (207.5,88.67) .. controls (198.76,88.67) and (191.67,81.58) .. (191.67,72.83) -- cycle ;
\draw  [draw opacity=0][fill={rgb, 255:red, 74; green, 144; blue, 226 }  ,fill opacity=1 ] (240.67,72.5) .. controls (240.67,63.76) and (247.76,56.67) .. (256.5,56.67) .. controls (265.24,56.67) and (272.33,63.76) .. (272.33,72.5) .. controls (272.33,81.24) and (265.24,88.33) .. (256.5,88.33) .. controls (247.76,88.33) and (240.67,81.24) .. (240.67,72.5) -- cycle ;
\draw  [draw opacity=0][fill={rgb, 255:red, 74; green, 144; blue, 226 }  ,fill opacity=1 ] (216,32.83) .. controls (216,24.09) and (223.09,17) .. (231.83,17) .. controls (240.58,17) and (247.67,24.09) .. (247.67,32.83) .. controls (247.67,41.58) and (240.58,48.67) .. (231.83,48.67) .. controls (223.09,48.67) and (216,41.58) .. (216,32.83) -- cycle ;
\draw [color={rgb, 255:red, 74; green, 144; blue, 226 }  ,draw opacity=1 ][line width=1.5]    (231.83,32.83) -- (256.5,72.5) ;
\draw [color={rgb, 255:red, 74; green, 144; blue, 226 }  ,draw opacity=1 ][line width=1.5]    (232.83,128.5) -- (208.5,168.5) ;
\draw [color={rgb, 255:red, 74; green, 144; blue, 226 }  ,draw opacity=1 ][line width=1.5]    (208.5,264.17) -- (257.5,263.83) ;
\draw [color={rgb, 255:red, 74; green, 144; blue, 226 }  ,draw opacity=1 ][line width=1.5]    (446.3,126.9) -- (470.97,166.57) ;
\draw [color={rgb, 255:red, 74; green, 144; blue, 226 }  ,draw opacity=1 ][line width=1.5]    (446.3,126.9) -- (421.97,166.9) ;
\draw  [draw opacity=0][fill={rgb, 255:red, 74; green, 144; blue, 226 }  ,fill opacity=1 ] (404.87,77.5) .. controls (404.87,68.76) and (411.96,61.67) .. (420.7,61.67) .. controls (429.44,61.67) and (436.53,68.76) .. (436.53,77.5) .. controls (436.53,86.24) and (429.44,93.33) .. (420.7,93.33) .. controls (411.96,93.33) and (404.87,86.24) .. (404.87,77.5) -- cycle ;
\draw  [draw opacity=0][fill={rgb, 255:red, 74; green, 144; blue, 226 }  ,fill opacity=1 ] (453.87,77.17) .. controls (453.87,68.42) and (460.96,61.33) .. (469.7,61.33) .. controls (478.44,61.33) and (485.53,68.42) .. (485.53,77.17) .. controls (485.53,85.91) and (478.44,93) .. (469.7,93) .. controls (460.96,93) and (453.87,85.91) .. (453.87,77.17) -- cycle ;
\draw  [draw opacity=0][fill={rgb, 255:red, 74; green, 144; blue, 226 }  ,fill opacity=1 ] (429.2,37.5) .. controls (429.2,28.76) and (436.29,21.67) .. (445.03,21.67) .. controls (453.78,21.67) and (460.87,28.76) .. (460.87,37.5) .. controls (460.87,46.24) and (453.78,53.33) .. (445.03,53.33) .. controls (436.29,53.33) and (429.2,46.24) .. (429.2,37.5) -- cycle ;
\draw [color={rgb, 255:red, 74; green, 144; blue, 226 }  ,draw opacity=1 ][line width=1.5]    (420.7,77.5) -- (469.7,77.17) ;
\draw [color={rgb, 255:red, 74; green, 144; blue, 226 }  ,draw opacity=1 ][line width=1.5]    (445.03,37.5) -- (420.7,77.5) ;
\draw  [draw opacity=0][fill={rgb, 255:red, 80; green, 227; blue, 194 }  ,fill opacity=0.3 ] (527.67,152.33) .. controls (527.67,127.48) and (547.81,107.33) .. (572.67,107.33) -- (602.67,107.33) .. controls (627.52,107.33) and (647.67,127.48) .. (647.67,152.33) -- (647.67,152.33) .. controls (647.67,177.19) and (627.52,197.33) .. (602.67,197.33) -- (572.67,197.33) .. controls (547.81,197.33) and (527.67,177.19) .. (527.67,152.33) -- cycle ;
\draw  [draw opacity=0][fill={rgb, 255:red, 74; green, 144; blue, 226 }  ,fill opacity=1 ] (547.33,170.5) .. controls (547.33,161.76) and (554.42,154.67) .. (563.17,154.67) .. controls (571.91,154.67) and (579,161.76) .. (579,170.5) .. controls (579,179.24) and (571.91,186.33) .. (563.17,186.33) .. controls (554.42,186.33) and (547.33,179.24) .. (547.33,170.5) -- cycle ;
\draw  [draw opacity=0][fill={rgb, 255:red, 74; green, 144; blue, 226 }  ,fill opacity=1 ] (596.33,170.17) .. controls (596.33,161.42) and (603.42,154.33) .. (612.17,154.33) .. controls (620.91,154.33) and (628,161.42) .. (628,170.17) .. controls (628,178.91) and (620.91,186) .. (612.17,186) .. controls (603.42,186) and (596.33,178.91) .. (596.33,170.17) -- cycle ;
\draw  [draw opacity=0][fill={rgb, 255:red, 74; green, 144; blue, 226 }  ,fill opacity=1 ] (571.67,130.5) .. controls (571.67,121.76) and (578.76,114.67) .. (587.5,114.67) .. controls (596.24,114.67) and (603.33,121.76) .. (603.33,130.5) .. controls (603.33,139.24) and (596.24,146.33) .. (587.5,146.33) .. controls (578.76,146.33) and (571.67,139.24) .. (571.67,130.5) -- cycle ;
\draw  [color={rgb, 255:red, 74; green, 144; blue, 226 }  ,draw opacity=1 ][line width=1.5]  (587.67,130.5) -- (612.17,170.17) -- (563.17,170.17) -- cycle ;
\draw  [draw opacity=0][fill={rgb, 255:red, 74; green, 144; blue, 226 }  ,fill opacity=1 ] (406.13,256.57) .. controls (406.13,247.82) and (413.22,240.73) .. (421.97,240.73) .. controls (430.71,240.73) and (437.8,247.82) .. (437.8,256.57) .. controls (437.8,265.31) and (430.71,272.4) .. (421.97,272.4) .. controls (413.22,272.4) and (406.13,265.31) .. (406.13,256.57) -- cycle ;
\draw  [draw opacity=0][fill={rgb, 255:red, 74; green, 144; blue, 226 }  ,fill opacity=1 ] (455.13,256.23) .. controls (455.13,247.49) and (462.22,240.4) .. (470.97,240.4) .. controls (479.71,240.4) and (486.8,247.49) .. (486.8,256.23) .. controls (486.8,264.98) and (479.71,272.07) .. (470.97,272.07) .. controls (462.22,272.07) and (455.13,264.98) .. (455.13,256.23) -- cycle ;
\draw  [draw opacity=0][fill={rgb, 255:red, 74; green, 144; blue, 226 }  ,fill opacity=1 ] (430.47,216.57) .. controls (430.47,207.82) and (437.56,200.73) .. (446.3,200.73) .. controls (455.04,200.73) and (462.13,207.82) .. (462.13,216.57) .. controls (462.13,225.31) and (455.04,232.4) .. (446.3,232.4) .. controls (437.56,232.4) and (430.47,225.31) .. (430.47,216.57) -- cycle ;
\draw [color={rgb, 255:red, 74; green, 144; blue, 226 }  ,draw opacity=1 ][line width=1.5]    (446.3,216.57) -- (470.97,256.23) ;
\draw [color={rgb, 255:red, 74; green, 144; blue, 226 }  ,draw opacity=1 ][line width=1.5]    (421.97,256.57) -- (470.97,256.23) ;
\draw [color={rgb, 255:red, 184; green, 233; blue, 134 }  ,draw opacity=0.3 ]   (499.88,39.25) -- (505.88,33.75) ;

\draw (493.33,303.31) node [anchor=north west][inner sep=0.75pt]   [align=left] {(iii)};
\draw (209,302.65) node [anchor=north west][inner sep=0.75pt]   [align=left] {(ii)};
\draw (33,207.3) node [anchor=north west][inner sep=0.75pt]   [align=left] {(i)};

\end{tikzpicture}

    \caption{All the five LC classes of the eight 3 qubit graph states, corresponding to the separable class (i), the class corresponding to a Bell pair with a separated qubit (ii), and the genuinely entangled class (iii), containing graphs with two and three edges ($\ket{W}$ and $\ket{\text{GHZ}}$ states, respectively).}
    \label{fig:graphexamples}
\end{figure}
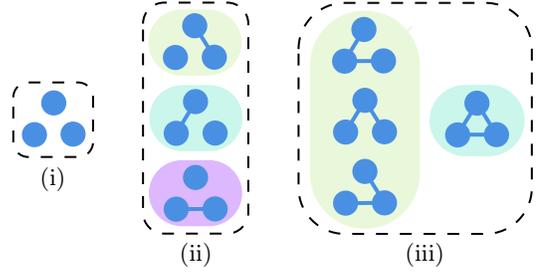

\prlsection{Beyond two parties}{sec:BeyondTwoParties} We consider now the multipartite scenario. Specifically, to illustrate our framework~\cite{supp}, we will focus on the tripartite scenario and on the W-state \cite{dur2000three} of the form $(1/\sqrt{3})(\ket{001}+\ket{010}+\ket{100})$. As discussed above, any Bell inequality respected by all possible graph-states will define a valid magic witness. As illustrated in Fig. \ref{fig:graphexamples}, the tripartite scenario has 8 different graph-states, that up to local unitaries and permutation of the parties can be subsumed by only three classes of entangled states: i) $\ket{000}$ (fully separable), ii) $\ket{\Phi^{+}}\ket{0}$ (2-separable), iii) $(1/\sqrt{2})\ket{000}+\ket{111}$ (genuinely tripartite entangled).

As shown in the Supplemental Material \cite{supp}, the inequality $S_3+R_2 \leq 6$~\cite{PhysRevLett.108.110501} is respected by all stabilizer states. The term $R_2$ involves only two-body expectation values as
\begin{align}
    R_2=\sum_{i=0,1}\braket{A_1^{i}A_2^{i\oplus1}  + A_1^{i}A_3^{i\oplus1}  +A_2^{i}A_3^{i\oplus1} }, \label{eq:R2}
\end{align}
and $S_3$ is the Svetlichny polynomial \cite{PhysRevD.35.3066} 
\begin{equation}
    S_3 = \left\langle \sum_{i=0,1}(-1)^iA_1^iA_2^iA_3^i +\sum_{cyc_1}A_1^iA_2^jA_3^k -\sum_{cyc_2}A_1^iA_2^jA_3^k \right\rangle, \label{eq: svetlichny}
\end{equation}
where the symbol $\oplus$ denotes sum modulo 2 and $cyc_n$ denotes all values of $i,j,k \in \{0,1\}$ subject to $i+j+k =n$. The W-state can surpass this bound reaching up to $S_3+R_2=7.26$~\cite{PhysRevLett.108.110501}, a violation that witnesses the non-stabilizerness of this state.

\prlsection{Discussion}{sec:Discussion} In this work, we have demonstrated how Bell inequalities can serve as effective tools for certifying non-stabilizerness in quantum states. While Bell nonlocality and magic arise from distinct aspects of quantum theory, we have shown that tailored Bell inequalities can reveal the presence of magic, bridging the gap between these two fundamental concepts. This insight provides a novel approach to characterizing quantum resources, particularly in the context of quantum computation, where non-stabilizerness is a key ingredient for quantum advantage.

Our analysis focuses on simple Bell scenarios, leveraging few-party and few-qudit configurations to identify regions of the correlation space where magic manifests. Unlike complex many-body systems—where stabilizer polytope characterization is computationally intractable—our approach provides a tractable way to witness non-stabilizerness through Bell inequalities. In particular, we establish the emergence of magic in the simplest Bell scenario via the tilted CHSH inequality, showing how its tilting parameter can be exploited to detect nonlocal correlations stemming from magic states using quadratically constrained quadratic programming (QCQP). Extending this framework, we explore higher-dimensional and multipartite scenarios, demonstrating that CGLMP inequalities naturally act as magic witnesses in the former, while generalized Svetlichny-type inequalities offer a pathway for multipartite certification, particularly in the case of non-graph states.

While the Bell inequalities presented here are sufficient to reveal non-stabilizerness, their optimality for this task remains an open question. Future work may explore whether more refined inequalities could provide stronger or more robust certification of magic, particularly in systems of increasing dimension and number of parties. Given the central role of non-stabilizerness in quantum computation, further investigation into its connection with Bell nonlocality may yield deeper insights into the fundamental resources enabling quantum advantage.

\section*{Acknowledgements}
We thank Pedro Lauand for helpful discussions. This work was supported by the Simons Foundation (Grant Number 1023171, RC), the Brazilian National Council for Scientific and Technological Development (CNPq, Grants No.307295/2020-6 and No.403181/2024-0), the Financiadora de Estudos e Projetos (grant 1699/24 IIF-FINEP), the Brazilian Coordination of Superior Level Staff Improvement (CAPES), Instituto Serrapilheira (Chamada 2020) and the Austrian Science Fund (FWF) through the Elise Richter project V1037. DP acknowledges funding from MUR PRIN (Project 2022SW3RPY).

\bibliography{main.bib}

\end{document}


\title{Supplemental Material: Witnessing Magic with Bell Inequalities}

\author{Rafael A. Mac\^{e}do}
\affiliation{Departamento de F\'isica Te\'orica e Experimental, Universidade Federal do Rio Grande do Norte, 59078-970 Natal-RN, Brazil}
\affiliation{International Institute of Physics, Federal University of Rio Grande do Norte, 59078-970, Natal, Brazil}
\author{Patrick Andriolo}
\affiliation{International Institute of Physics, Federal University of Rio Grande do Norte, 59078-970, Natal, Brazil}
\affiliation{Physics Institute, University of São Paulo, Rua do Matão, 1371, São Paulo, Brazil}
\affiliation{Atominstitut, Technische Universität Wien, Stadionallee 2, 1020 Vienna, Austria}
\author{Santiago Zamora}
\affiliation{Departamento de F\'isica Te\'orica e Experimental, Universidade Federal do Rio Grande do Norte, 59078-970 Natal-RN, Brazil}
\affiliation{International Institute of Physics, Federal University of Rio Grande do Norte, 59078-970, Natal, Brazil}
\author{Davide Poderini}
\affiliation{International Institute of Physics, Federal University of Rio Grande do Norte, 59078-970, Natal, Brazil}
\affiliation{Universit\'a degli Studi di Pavia, Dipartimento di Fisica, QUIT Group, via Bassi 6, 27100 Pavia, Italy}
\author{Rafael Chaves}
\affiliation{International Institute of Physics, Federal University of Rio Grande do Norte, 59078-970, Natal, Brazil}
\affiliation{School of Science and Technology, Federal University of Rio Grande do Norte, 59078-970 Natal, Brazil}

\begin{abstract}
    On this supplemental material we provide the details on the construction of stabilizer scenarios for generic Bell inequalities, we discuss about the complexity of the problem and we compute the stabilizer value of a three-party Bell inequality.
\end{abstract}
\maketitle

\section{Stabilizer scenario for generic Bell inequalities}
In this section we develop our framework for multi-qudit systems. We define what we call the \textit{stabilizer scenario}, framing it as a classical optimization problem and presenting our main result in  Theorem 1.

Let us begin by considering finite-dimensional quantum correlations. On an $n$-party Bell scenario, this implies that we  restrict ourselves to a Hilbert space of the form
$\mathcal H = \otimes_{i=1}^n\mathbb C^{d_i}$  where it is assumed the maximal tensor product decomposition: every individual system $i$ has prime dimension $d_i$.   Assume now, that each party $j$ can perform $m_j$ distinct measurements, and each measurement can result in 1 of $d_j$ outcomes. We restrict to projectors and we denote them by $M^{(j)}_{a_j|x_j}$ where $x_j \in \{0,1,\cdots, m_j-1\}$ denotes the input  and $a_j \in \{0,1,\cdots, d_j-1\}$ denotes the outcome.  Given a quantum state $\rho \in \mathcal D(\mathcal H)$,  by each party measuring its subsystem the following behavior is generated
\begin{equation}
    p(a_1, a_2, \cdots, a_n| x_1, x_2, \cdots, x_n) = \mathrm{tr}(\rho M^{(1)}_{a_1|x_1}\otimes M^{(2)}_{a_2|x_2}\otimes\cdots \otimes M^{(n)}_{a_n|x_n})\;.
\end{equation}
  Let $[I^\mathbf a_\mathbf x]$ be $\prod_{i=1}^nd_i m_i$ numbers defining the inequality $I =\sum_{\mathbf a, \mathbf x}I^\mathbf a_\mathbf x\; p(\mathbf a|\mathbf x)$, with $\mathbf a= (a_1, a_2, \cdots, a_n)$ and $\mathbf x = (x_1, x_2, \cdots, x_n)$. The corresponding maximal quantum value is given by the maximization problem
\begin{align}
I_Q(\mathcal H) \equiv &\max_{(M,\psi \in \mathcal H)} \sum_{\mathbf a, \mathbf x} I^{\mathbf a}_\mathbf x \;\langle \psi| \bigotimes_{i=1}^n M^{(i)}_{a_i|x_i}|\psi\rangle \;, \label{eq:genericineq}\\
&M^{(i)}_{a_i|x_i} M^{(i)}_{a'_i |x_i}= \delta_{a'_i , a_i} M^{(i)}_{a_i|x_i} \quad \forall i,x_i\;, \label{eq:measurementcond1}\\
&\sum_{a=0}^{d_i-1} M^{(i)}_{a|x_i} = 1\quad \forall i,x_i\;, \label{eq:measurementcond2}
\end{align}
where we are optimizing over the projectors $M_{\mathbf a|\mathbf x} = \otimes_{i}M^{(i)}_{a_i|x_i}$ and the states $|\psi\rangle \in \mathcal H$, which by convexity of the quantum states and linearity of the Bell inequalities we can restrict to pure states. Note that this is not the case in general regarding the measurements. Here we considered only projectors, but for high dimensional systems this is not a trivial assumption.   Also note that the problem is restricted to a specific Hilbert space, so the quantum bound obtained by the maximization~(\ref{eq:genericineq}) may not be the true maximum quantum value obtained when the optimization is carried on over all Hilbert spaces.
 We now establish our first result.
\\
\begin{lemma}
Consider $I_Q(\mathcal H)$ as defined on Eq.~(\ref{eq:genericineq}). Then, the optimization over  projectors is equivalent to the following optimization over unitary operators:
\begin{align}
I_Q(\mathcal H) = &\max_{(U,\psi \in \mathcal H)} \sum_{\mathbf k, \mathbf x} \tilde{I}^{\mathbf k}_\mathbf x \;\langle \psi| \bigotimes_{i=1}^n (U^{x_i}_{i})^{k_i}|\psi\rangle \;, \label{eq:genericunitaryineq}\\
&U^{x_i\dagger}_{i}U^{x_i}_{i}= 1 \quad \forall i,x_i\;, \label{eq:unitarycond1}\\
&(U^{x_i}_{i})^{d_i} = 1\quad \forall i,x_i\;, \label{eq:unitarycond2}
\end{align}
where
\begin{equation}
\tilde I^\mathbf k_\mathbf x = \frac{1}{\prod_{i=1}^nd_i} \sum_{\mathbf a \in \bigoplus_{i=1}^n \mathbb F_{d_i}} \exp \left[2 \pi i \sum_{j=1}^n \frac {k_i a_i}{d_i}\right] I ^\mathbf a_\mathbf x\;,\label{eq: FourierTransform_coeff}
\end{equation}
is the Fourier transform of the coefficients $[I^\mathbf a_\mathbf x]$, with $\mathbf k \in \bigoplus_{i=1}^n \mathbb F_{d_i}$ an element of the direct sum of the $n$ Galois Fields $\mathbb F_{d_i}$.
\end{lemma}

\begin{proof}
 First, let us define the discrete Fourier transform of the behaviors as
\begin{equation}
C^\mathbf k_\mathbf x = \sum_{\mathbf a} e^{2\pi i f_\mathbf d(\mathbf k , \mathbf a)} p(\mathbf a | \mathbf x) =  \sum_{\mathbf a} e^{2\pi i f_\mathbf d(\mathbf k , \mathbf a)} \langle \psi| \bigotimes_{i=1}^n M^{(i)}_{a_i|x_i}|\psi\rangle\;,
\end{equation}
where $f_\mathbf d(\mathbf k , \mathbf a)= \sum_{j=1}^n k_j a_j/d_j$. From the properties of the discrete Fourier transform it follows that,
\begin{equation}
    \sum_{\mathbf k, \mathbf x }\tilde{I}^{\mathbf k}_\mathbf x\; C^\mathbf k_\mathbf x = \sum_{\mathbf a , \mathbf x}I^{\mathbf a}_\mathbf x \;p(\mathbf a |\mathbf x)\;,
\end{equation}
where $\tilde{I}^{\mathbf k}_\mathbf x$ is defined in Eq.~(\ref{eq: FourierTransform_coeff}). Now, define:
\begin{equation}
U^{x_i}_{i} \equiv \sum_{a_i =0}^{d_i-1}e^{2\pi i  a_i/d_i}M^{(i)}_{a_i|x_i}\;,
\label{eq:unitarydecomp}
\end{equation}
noting that $C^\mathbf k_\mathbf x = \langle \psi|\bigotimes_{i=1}^n (U^{x_i}_{i})^{k_i}|\psi\rangle$.  Then, the orthogonality and completeness of the projectors $\{M^{(i)}_{a_i|x_i}\}$, implies that the operators $U^{x_i}_{ i}$ are unitary and satisfy $(U^{x_i}_{ i})^{d_i} = 1$. This proves the necessity of Eqs. (\ref{eq:measurementcond1}, \ref{eq:measurementcond2}). They are also sufficient: Assume that the unitaries have a spectral decomposition
\begin{equation}
U^{x_i}_{i} = \sum_{a=0}^{d_i-1}e^{i\lambda_a}M_{a|x_i}^{(i)}\;,
\end{equation}
where $\{M_{a|x_i}^{(i)}\}$ are a set of projectors satisfying Eqs. (\ref{eq:measurementcond1}
, \ref{eq:measurementcond2}). Then, from $(U^{x_i}_{i})^{d_i}=1$, it follows that their eigenvalues are the $d$-th roots of unity, $i.e.$ they have the form of Eq.~(\ref{eq:unitarydecomp}).
Altogether, it shows that the optimization over projectors~(\ref{eq:genericineq}) is equivalent to the optimization over unitaries~(\ref{eq:genericunitaryineq}).
\end{proof}

\noindent 
To simplify the notation, let us denote
\begin{equation}
    \mathrm U_\mathbf{d}   \equiv \left\{\otimes_{i=1}^n U_{i}\;|\;U_i\in U(\mathbb C^{{d_i}})\;,\; (U_i)^{d_i}=1 \right\}\;,
    \label{eq:LUset}
\end{equation}
as the group of local unitaries satisfying Eqs.~(\ref{eq:unitarycond1}, \ref{eq:unitarycond2}). Then, by defining $U^\mathbf k_\mathbf x = \bigotimes_{i=1}^n (U^{x_i}_{i})^{k_i}$, we can write
\begin{equation}
I_\mathrm{Q}(\mathcal H) = \max_{(U \in \mathrm U_\mathbf d,\psi \in \mathcal H)} \sum_{\mathbf k, \mathbf x} \tilde{I}^{\mathbf k}_\mathbf x \;\langle \psi|U^\mathbf k_\mathbf x|\psi\rangle \;,
\end{equation}
\noindent
where the optimization considers all $\sum_i m_i$ unitaries $ U_\mathbf x = \otimes_{i=1}^n U^{x_i}_{i} \in \mathrm U_\mathbf d$.

Now, with the aim of defining the stabilizer scenario, let us first introduce the stabilizer states in $\mathcal H = \otimes_{i=1}^n \mathbb C^{d_i}$.  Consider the $1$ qudit Pauli group $\mathcal P_d$ with prime dimension $d$. Its elements are  known as displacement operators defined as
\begin{equation}
    D_{\mathbf a } = \omega_d^{-\frac{a_1  a_2}{2}} X^{ a_1}Z^{ a_2} , \quad \quad  \mathbf a = ( a_1,a_2) \in \mathbb F_d \times \mathbb F_d\;,
\end{equation}
where $\omega_2= i$ and $\omega_{d \neq 2}  = e^{2\pi i/d}$. The operators $X$ and $Z$ are the generalized Pauli operators which satisfy the algebra $X^d=Z^d=1$ and $XZ = \omega_d ZX$. The group law is defined by $D_{\mathbf a }D_{\mathbf b}  = \omega^{S(\mathbf a, \mathbf b)/2}_d D_{\mathbf a + \mathbf b}$, where $S(\mathbf a, \mathbf b) = a_1 b_2-a_2b_1$ is the symplectic inner product. 
Given a Hilbert space of dimension $d$ spanned by a basis $\{\ket{i}\}^{d-1}_{i=0}$, the generalized Pauli operators have an irreducible representation $\rho_d: \mathcal P_d\to U(\mathbb C^d)$ defined by
\begin{equation}
    \rho_d(X) = \sum_{a=0}^{d-1} (|a+1\rangle \langle a|+\mathrm{h.c}) \quad ; \quad \rho_d(Z) = \sum_{a=0}^{d-1}\omega_d^a|a\rangle \langle a|\;,
\end{equation}
which generate a basis of the space of complex matrices  with elements given by products of the form
\begin{align}
    \rho_d(X)^k \rho_d(Z)^j =\sum_{a=0}^{d-1} \omega^{aj}_d \ket{a+k}\bra{a} ,
\end{align}
where  $k,j\in \mathbb F_d$. Hence, given a list of local dimensions $\mathbf d  = (d_1, d_2,\cdots, d_n)$, the corresponding Pauli group is defined as $\mathcal P_\mathbf d \equiv \mathcal P_{d_1}\times \mathcal P_{d_2}\times \cdots \times \mathcal P_{d_n}$ and when restricting to the fixed Hilbert space $\mathcal H = \bigotimes_{i=1}^n \mathbb C^{d_i}$ its elements have the tensor product representation $\rho_{d_1}\otimes \rho_{d_2}\otimes \cdots \otimes \rho_{d_n}$, implicitly assumed from now on.

A stabilizer group \cite{gottesman2016surviving} is defined by an abelian subgroup $\mathcal S \subseteq \mathcal P_\mathbf d$, such that \textit{i)} $e^{i\phi} \notin \mathcal S$ for all $\phi \neq 0$ and \textit{ii)} $\mathcal S$ has $n$ generators. The corresponding stabilizer state $|\mathcal S\rangle \in \mathcal H$ is the state stabilized by the elements of the group, \textit{i.e.} $s|\mathcal S\rangle = |\mathcal S\rangle$ for all $ s \in \mathcal S$. Since the correspondence between the subgroups and the stabilizer states is one to one, we will denote with $\mathrm{STAB}_\mathbf d$ both  the set of all subgroups $\{\mathcal{S}\}$ and  the set of all stabilizer states $\{|\mathcal S\rangle\}$. 

The Pauli group has as normalizer the Clifford group, \textit{i.e.} the Clifford group is defined as the subset of all the unitaries $\mathcal C \ell_\mathbf d \subseteq U(\mathbb C^d)$ that sends Pauli operators into Pauli operators by conjugation. That is, given some $C \in \mathcal C\ell_\mathbf d$, for all $D \in \mathcal P_\mathbf d$:
\begin{equation}
    \exists D^\prime \in \mathcal P_\mathbf d \;:\quad C D C^\dagger =e^{i \phi (D, D^\prime)}D^\prime\;,
\end{equation}
where $\phi(D, D^\prime) \in [0,2\pi)$ is a phase depending on $D$ and $D'$. We can then define the $\mathbf d = (d_1, d_2, \cdots, d_n)$ stabilizer scenario of the inequality defined by the coefficients $[I^\mathbf a _\mathbf x]$ as the one where the states are restricted to $\mathrm{STAB}_\mathbf d$:
\begin{align}
I_\mathrm{STAB}(\mathcal H, \mathbf d) \equiv &\max_{(U \in \mathrm U_\mathbf d,\mathcal S \in \mathrm{STAB}_\mathbf d)} \sum_{\mathbf k, \mathbf x} \tilde{I}^{\mathbf k}_\mathbf x \;\langle \mathcal S| U^\mathbf k_\mathbf x|\mathcal S\rangle \;, 
\end{align}
where we used Lemma 1 to define the optimization over all unitaries.

Now we will show that the above optimization can be further simplified. First, by considering the local-unitary invariance of the Bell inequalities, we can reduce the optimization to representatives of specific classes of stabilizer states as stated in the following lemma.

\begin{lemma}
The value of the $\mathbf d=(d_1, d_2, \cdots, d_n)$ stabilizer scenario of an inequality is obtained by 
optimizing over \textbf{L}ocal \textbf{C}lifford classes (LC) of $\mathrm{STAB}_{\mathbf d}$:
\begin{align}
I_\mathrm{STAB}(\mathcal H, \mathbf d) = &\max_{(U \in \mathrm U_\mathbf d,\mathcal S \in [\mathrm{STAB}_\mathbf d]_\mathrm{LC})} \sum_{\mathbf k, \mathbf x} \tilde{I}^{\mathbf k}_\mathbf x \;\langle \mathcal S| U^\mathbf k_\mathbf x|\mathcal S\rangle \;, 
\label{eq:staboptform}
\end{align}
where the set of equivalence classes is defined as follows. Given $|\psi_1\rangle, |\psi_2\rangle \in \mathcal H$, 
\begin{equation}
|\psi_1\rangle \overset{\text{LC}}{\sim} |\psi_2\rangle \;\Leftrightarrow\; \exists C \in \mathcal C\ell_{d_1}\otimes \mathcal C\ell_{d_2}\otimes \cdots \otimes \mathcal C \ell_{d_n} \;,\; |\psi_1\rangle = C |\psi_2\rangle.
\end{equation}
\end{lemma}

\begin{proof}
Consider the state $|\psi\rangle \in \mathcal H$  such that it satisfies  
\begin{equation}
I_Q(|\psi\rangle) \equiv \max_{V \in \mathrm U_\mathbf d}\sum_{\mathbf k, \mathbf x} \tilde{I}^{\mathbf k}_\mathbf x \;\langle \mathcal \psi| V^\mathbf k_\mathbf x|\psi\rangle\;,
\end{equation}
for some Bell inequality $I_Q$. Note that for any set of local unitaries $\{U_i \in U(\mathbb C^{d_i})\}_i$ we get
\begin{equation}
I(|\psi\rangle) = I(U_1 \otimes U_2 \otimes \cdots \otimes U_n|\psi\rangle)\;,
\end{equation}
 since for all  $V \in \mathrm{U}_\mathbf d$, $[\otimes_i U_i]^\dagger V [\otimes_i U_i] \in \mathrm{U}_\mathbf d$. Hence, the optimization over states can be reduced to the optimization over only one representative of each  $\mathrm{LU}$ equivalence class of states,
\begin{align}
I(\mathcal H) = \max_{(U \in \mathrm U_\mathbf d,\psi \in [\mathcal H]_\mathrm{LU})} \sum_{\mathbf k, \mathbf x} \tilde{I}^{\mathbf k}_\mathbf x \;\langle \psi|U^\mathbf k_\mathbf x|\psi\rangle \;.
\end{align}
This is the well-known local-unitary invariance of the Bell inequalities. However, in the stabilizer scenario, we restrict the states to lie in $\mathrm{STAB}_\mathbf d$. This implies that the optimization over Local Unitary (LU) equivalence classes of states can be reduced to the optimization over LC equivalence classes of stabilizer states, since those are the LU transformations that maps stabilizer states into themselves.
\end{proof}
 
We turn now our attention to the structure of the LC equivalence classes of stabilizer states. It turns out that the stabilizer states $\ket{\mathcal S} \in \mathrm{STAB}_\mathbf d $ can only  be a product state if the dimensions in $\mathbf d$ are coprime and distinct. We state this formally in Lemma \ref{Lemma: Stab_fact}. 
\begin{lemma}
\label{Lemma: Stab_fact}
 Let $\mathrm{STAB}_{\mathbf d = (d_1, d_2, \cdots, d_n)}$ be the set of stabilizer states on $\mathcal H = \bigotimes_{i=1}^n \mathbb C^{d_i}$, with all $d_i$ prime. Let $[\mathbf d]$ be the grouping of the dimensions $d_i$ into unique values. This is, $[\mathbf d]$ is the set of values defined by the equivalence class $d_i \sim d_j \;\Leftrightarrow d_i =d_j$which allows for the Hilbert decomposition $\mathcal H \cong \bigotimes_{\mathbf x \in [\mathbf d]}\mathcal H_\mathbf x$. Then, given $\mathcal S \in \mathrm{STAB}_\mathbf d$, 
\begin{equation}
|\mathcal S\rangle = \bigotimes_{\mathbf x \in [\mathbf d]}|\mathcal S_\mathbf x\rangle\;,
\end{equation}
where $\mathcal S_\mathbf x \in \mathrm{STAB}_\mathbf x$ is a stabilizer group acting on $\mathcal H_\mathbf x$.    
\end{lemma}
 \begin{proof}
  The grouping into unique values of the dimensions induces a decomposition of the stabilizer group. The full Pauli group decomposes as $\mathcal P_\mathbf d = \bigtimes_{\mathbf x \in [\mathbf d]} \mathcal P_{d_\mathbf x}$, where $\mathcal P_{d_\mathbf x}$ acts non-trivially only on $\mathcal H_\mathbf x$. On the other hand, note that $|\mathcal S| = \mathrm{dim}(\mathcal H) = \prod_{i=1}^n d_i = \prod_{\mathbf x \in [\mathbf d]} d_\mathbf x^{|\mathbf x|}$, which corresponds to a prime decomposition of $|\mathcal S|$. The fundamental theorem of finite abelian groups \cite{lang2012algebra} states that there is a decomposition satisfying the prime decomposition:
\begin{equation}
\mathcal S \cong \bigtimes_{\mathbf x \in [\mathbf d]} \mathcal S_\mathbf x \;,
\end{equation}
where each of the groups $\mathcal S_{\mathbf x}$ has order $|\mathcal S_{\mathbf x}| \equiv D_{\mathbf{x}} = d_{\mathbf{x}}^{|\mathbf x|}$, and all its elements must have order $d_{\mathbf x}^l$, for some $l \in \mathbb Z_{>0}$. However, due to the coprimality of the dimensions, it follows that $\mathcal S_\mathbf y \subseteq \mathcal P_{d_\mathbf y}$, since that is the only factor in the decomposition $\bigtimes_{\mathbf x \in [\mathbf d]} \mathcal P_{d_\mathbf x}$ which contains elements with order as a power of $d_\mathbf x$. Hence, all elements $s \in \mathcal S_\mathbf x$ \textit{i)} have order $d_\mathbf x$ and \textit{ii)} have only non-trivial support at $\mathcal H_\mathbf x$, that is, they are of the form $s = s_\mathbf x \otimes 1_{\bar {\mathbf x}}$.

Now, we will review how the algebraic structure of the stabilizer group completely fixes the corresponding bipartite entanglement structure \cite{fattal2004entanglement}. Denoting $\{g_i\}_{i=1}^n$ as the generating set of the group $\mathcal S = \langle g_1, g_2,\cdots, g_n \rangle$ and  using the fact that $g_i \in \mathrm U_{d_i}$, we note that the projector into the $g_i=1$ subspace is given by $P_i = d_i^{-1} \sum_{k=0}^{d_i-1}g_i^k$. Hence,
\begin{equation}
|\mathcal S \rangle \langle \mathcal S| = \prod_{i=1}^nP_i = \prod_{i=1}^n \left(\frac{1}{d_i} \sum_{k_i=0}^{d_i-1} g_i^{k_i}\right)
 =  \frac{1}{D}\sum_{k_1, k_2, \cdots, k_n} \prod_{i=1}^{n}g_i^{k_i} = \frac{1}{D}\sum_{s \in \mathcal S} s\;,
\end{equation}
where on the third equality each $k_i \in [d_i-1] =\{0,1,\cdots, d_i-1\}$, and we have used on the last equality the fact that every $s$ can be written as a product of the generators. Given some $\mathbf x \in [\mathbf d]$, there is a decomposition $\mathcal H = \mathcal H_ \mathbf x \otimes \mathcal H_{\bar{\mathbf x}}$, with a corresponding reduced density matrix:
\begin{equation}
    \rho_\mathbf x = \mathrm{tr}_{\bar{\mathbf x}}|\mathcal S \rangle \langle \mathcal S| = \frac{1}{D} \sum_{s \in \mathcal S} \mathrm{tr}_{\bar{\mathbf x}}(s)\;\label{eq: rho_x}.
\end{equation}
Note that since the stabilizers are  Pauli operators with zero trace, it implies that if $\mathrm{tr}_{\bar {\mathbf x}}(s) \neq 0$, then $s \in \mathcal S_\mathbf x$. Therefore $s = s_\mathbf x \otimes 1_{\bar {\mathbf x}}$ and $\mathrm{tr}_{\bar {\mathbf x}}(s) =D_{\bar{\mathbf x} } s_\mathbf x$ allowing us to rewrite  Eq.~(\ref{eq: rho_x}) as
\begin{equation}
    \rho_\mathbf x = \frac{1}{D_\mathbf x} \sum_{s_\mathbf x \otimes 1_{\bar{\mathbf x }} \in \mathcal S_\mathbf x}s_\mathbf x\;.
\end{equation}
From this decomposition, it follows that $\mathrm{tr}(\rho^2_\mathbf x) = |\mathcal S_\mathbf x|/D_\mathbf x =1$, showing that the purity of any reduced matrix in $\mathbf x \in [\mathbf d]$ is 1, showing the Lemma.    
 \end{proof}

 The above Lemma shows that quantum distributions obtained by applying measurements on each subsystem of a multipartite stabilizer state factorize if the subsystems are in Hilbert spaces of different dimensions, since the associated state is a product state. In particular,  for a Bell scenario, it means that they can only generate local correlations. Formally, given any inequality with local bound
\begin{align}
I_\mathrm{LOC} = &\max_{p(\mathbf a |\mathbf x) } \sum_{\mathbf a, \mathbf x } I^\mathbf a_\mathbf x\; p(\mathbf a |\mathbf x)\;,\\
\text{s.t.} \nonumber\\
&p(\mathbf a |\mathbf x) = \sum_\lambda p(\lambda)\;  \prod_{i=1}^n p(a_i|x_i, \lambda), 
\label{eq:locconstraint1}\\
&p(a_i|x_i, \lambda) \in [0,1]\; \quad \forall i \in \{1,2,\cdots,n \},\\
&\sum_{a_i=0}^{d_i-1}p(a_i|x_i, \lambda)=1\; \quad \forall i \in \{1,2,\cdots,n \},
\label{eq:locconstraint2}
\end{align}
since a product state $|\psi\rangle =|\psi_1\rangle \otimes |\psi_2\rangle \otimes \cdots \otimes |\psi_n\rangle$ can only correspond to an instance of a local correlation with factorizing behaviors as $p(\mathbf a |\mathbf x)  =\prod_{i=1}^n p(a_i|x_i)$, it follows that:

\begin{corollary}
     Let $[I^\mathbf a_\mathbf x]$ be an inequality which considers measurements of  $\mathbf d = (d_1, d_2, \cdots, d_n)$ outcomes, all prime and distinct. Then, the corresponding stabilizer scenario on $\mathcal H = \bigotimes_{i=1}^n \mathbb C^{d_i}$ can at most saturate the local bound of $[I^\mathbf a_\mathbf x]$, $I_\mathrm{STAB}(\mathcal H, \mathbf d) \leq I_\mathrm{LOC}$.
\end{corollary}

Therefore, non-classical correlations can only appear for stabilizer states with subsystems  in Hilbert spaces of the same dimension. As in the case of qubits, their entanglement structure can be associated to a qudit graph state, which we define in the following. 

For prime $d$, let $A \in \mathbb F_d^{n\times n}$ be a symmetric adjacency matrix with entries mod $d$, assumed to have vanishing diagonals. Start by defining the control phase gate $\mathrm{CP}$ acting on $\mathbb C^d\otimes \mathbb C^d$  as
\begin{equation}
\mathrm{CP} \equiv \sum_{j=0}^{d-1}|j\rangle \langle j |\otimes Z^j\;.
\end{equation}
It can be verified that $\mathrm{CP} \in \mathcal C\ell_{d}\otimes \mathcal C\ell_d$ is a Clifford gate. This gate defines the state $|G(A)\rangle \in (\mathbb C^{d})^{\otimes n}$,
\begin{equation}
|G(A)\rangle \equiv \left(\prod_{i=1}^n \prod_{j=i+1}^n \mathrm{CP}^{A_{ij}}_{ij}\right) \bigotimes_{i=1}^n |+\rangle\;,
\end{equation}
where $|+\rangle$ is the $+1$ eigenstate of $X$, and $\mathrm{CP}_{ij}$ is the control phase gate acting on  the qudits $i$ and $j$. We will refer to $G(A)$ as the corresponding undirected, no-loop graph generated by $A$. Its stabilizer group can be computed to be:
\begin{equation}
    \mathcal S_{G(A)} = \left\langle \left\{ X_i \left(\prod_{j=1}^n Z_j^{-A_{ij}}\right)\right\}_{i=1}^n\right \rangle\;.
\end{equation}
It was shown in \cite{bahramgiri2006graph} that every stabilizer state is locally related to some graph state:

\noindent 
\begin{lemma}
\label{lemma: Stab_Graph_relation}
    (Lemma 7 of \cite{bahramgiri2006graph}).   
    Let $\mathcal S \subseteq \mathrm{STAB}_{(d, d, \cdots, d)}$ be a stabilizer state on $(\mathbb C^{d})^{\otimes n}$. Then, it exists a graph state $|G(A)\rangle \in \mathrm{STAB}_{(d,d,\cdots, d)}$ and unitaries $C_i \in \mathcal C\ell_d$ such that
\begin{equation}
|\mathcal S\rangle = \bigotimes_{i=1}^n C_i |G(A)\rangle\;.
\end{equation}
\end{lemma}

Thus, by combining Lemma~\ref{Lemma: Stab_fact} and Lemma~\ref{lemma: Stab_Graph_relation}, we conclude that optimizing over the equivalence classes of stabilizer states is equivalent to optimizing over the equivalence classes of graph states, where both classes are defined under local Clifford equivalence. In Ref.~\cite{bahramgiri2006graph}  the authors study these equivalence classes of graph states and their result can be summarize as follows. 

First, let  $A \in \mathbb F_d^{n\times n}$ be a symmetric adjacency matrix with vanishing diagonals, and consider $v \in \{1,2,\cdots,n\}$ as one of the vertices of $G(A)$. Now define two maps acting on $A$. The first one is defined by
\begin{equation}
\mathrm M_{b,v} A \equiv I_{v,b}  A I_{v,b} \in \mathbb F_d^{n \times n} \quad; \quad I_{v,b} = \mathrm{diag}(1,1,\cdots,b,\cdots, 1)\;,
\label{eq:Mmove}
\end{equation}
where $b \neq 0 \in \mathbb F_d$ is placed on the $v$th entry in $I_{v,b}$. The second map $\mathrm{L}_{a,v}A \in \mathbb F_d^{n \times n}$ can defined  element-wise as
\begin{align}
    (\mathrm{L}_{a,v}A)_{ij} &= A_{ij} + a A_{vi}A_{vj} \quad ; \quad i\neq j\;, \label{eq:LC1}\\
    (\mathrm{L}_{a,v}A)_{jj} &= A_{jj} = 0 \;, \label{eq:LC2}
\end{align}
where  $a \in \mathbb F_d$.  

\begin{figure}[h!]
    \centering
    \includegraphics[width=0.5\linewidth]{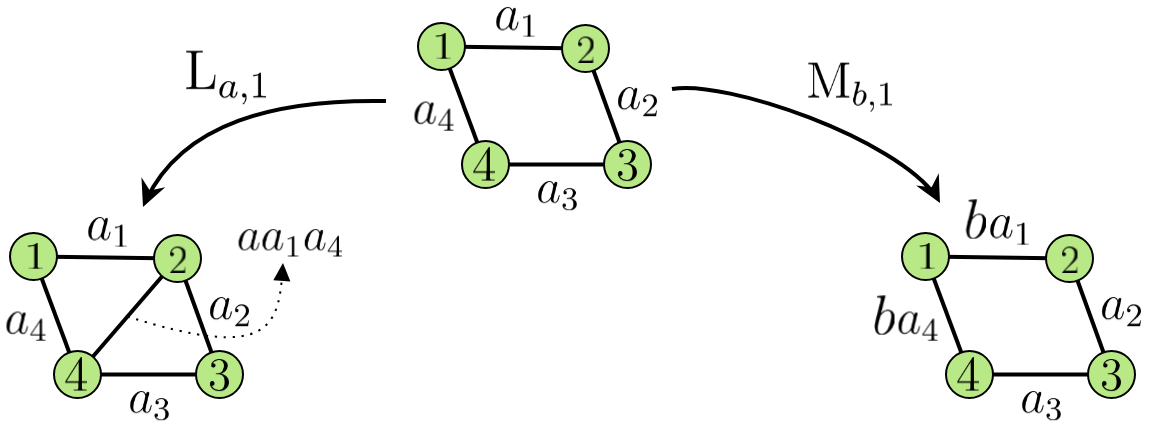}
    \caption{\textbf{Example of the action of $M$ and $L$.} Resulting graphs after the application of $\mathrm M$ and $L$ on $A$. Here $A$ is the adjacency matrix representing the four vertex graph with weights $\{a_1,a_2,a_3,a_4\}$ shown in the middle.}
    \label{fig:graphexamples}
\end{figure}

The last map is also referred as the \emph{local complementation} of a vertex $v$ in the binary $d=2$ case~\cite{van2004graphical}. See Fig.~\ref{fig:graphexamples} for an explicit example of the action of these maps. With these definitions the result of interest can be stated as in the following Lemma.\\

\begin{lemma}
\label{lemma: Equivalence_Graphs}
    (Theorem 5 of \cite{bahramgiri2006graph}) 
    Let $A, A^\prime  \in \mathbb  F_{d}^{n \times n}$ be two adjacency matrices with vanishing diagonals. Then,
\begin{equation}
    [|G(A)\rangle]_\mathrm{LC} = [|G(A^\prime)\rangle]_\mathrm{LC} \;\Leftrightarrow \;  A^\prime  = \mathrm{O}_{a_1,v_1}\circ \mathrm{O}_{a_2, v_2}\circ\cdots \mathrm{O}_{a_k, v_k}(A)\;,
\end{equation}
for some $v_1, v_2, \cdots, v_k \in \{1,2,\cdots, n\}$, $a_1, a_2, \cdots, a_k \in \mathbb F_d$, $\mathrm O \in \{\mathrm{M}, \mathrm{L}\}$ and $k>0$. That is, the adjacency matrices must be related by a composition of $\mathrm M$ and $\mathrm{L}$ operations.
\end{lemma}

This means that the problem of optimizing over Local Clifford classes reduces to finding all the adjacency matrices of graphs with $n$  vertices and maximum edge weight $d$ modulo $\mathrm M$ and $\mathrm{L}$ transformations. Under all the above considerations, we can now state our main result:

\begin{theorem}
\label{theo: main_result}
    Let $[A_n]_{\mathrm M, \mathrm L}$ denote the set of equivalence classes of adjacency matrices $ A \in \mathbb F_{d}^{n \times n}$, modulo $\mathrm M$ and $\mathrm L$ transformations. Then, the maximum value of a $\mathbf d= (d_1, d_2, \cdots, d_n)$ stabilizer scenario on $\mathcal H = \bigotimes_{i=1}^n \mathbb C^{d_i}$ is obtained by the following optimization problem:
\begin{align}
I_\mathrm{STAB}(\mathcal H, \mathbf d) = \max_{A \in \bigoplus_{\mathbf x \in [\mathbf d]}[A_{|\mathbf x|}]_{\mathrm M, \mathrm L}} &\max_{\{U_\mathbf x  = \bigotimes_i U^{x_i}_{i} \in (\mathbb C^{d_i \times d_i})^{\otimes n}\}}  \sum_{\mathbf k, \mathbf x}  \tilde I^{\mathbf k}_\mathbf x\; \langle G(A)| \bigotimes_{i=1}^n(U_{i}^{x_i})^{k_i}  |G(A)\rangle\;, \label{eq:thmopt}\\
& U_{i}^{x_i\dagger}U_{i}^{x_i} = 1 \quad \forall i,x_i, \\
& (U^{x_i}_{i})^{d_i} =1 \quad \forall i, x_i\;,
\end{align}
where the set   $ \bigoplus_{\mathbf x \in [\mathbf d]}[A_{|\mathbf x|}]_{\mathrm M, \mathrm L}$ contains all the direct sums of the representatives of the equivalence classes $[A_{|\mathbf x|}]_{\mathrm M, \mathrm L}$, this means that if   $ A \in \bigoplus_{\mathbf x \in [\mathbf d]}[A_{|\mathbf x|}]_{\mathrm M, \mathrm L}$, then  $A = \bigoplus_{\mathbf x \in [\mathbf d]} A_\mathbf x$, where $A_\mathbf x \in [A_{|\mathbf x|}]_{\mathrm M, \mathrm L}$ is a representative of an $\{\mathrm M, \mathrm L\}$ orbit. 
\end{theorem}

\begin{proof}
    The theorem follows if we show that we can map the optimization over stabilizer states to an optimization over adjacency matrices.

    Consider the problem of optimizing $\mathcal S \in [\mathrm{STAB}_\mathbf d]_\mathrm{LC}$. If we denote the set of graph states in vertices of homogeneous dimensions $\mathbf d^\prime = (d^\prime, d^\prime, \cdots, d^\prime)$ as $\mathrm{GRAPH}_{\mathbf d^\prime}$, we can change the optimization of stabilizer classes to an optimization over classes of graph states  clusters,
\begin{equation}
\mathcal S \in [\mathrm{STAB}_\mathbf d]_\mathrm{LC} \Leftrightarrow \mathcal S  = G(A) \;:\; G(A)\in \bigotimes_{\mathbf x \in [\mathbf d]}[\mathrm{GRAPH}_\mathbf x]_\mathrm{LC}\;,
\label{eq:stabtograph}
\end{equation}
since every stabilizer state can be written as a product of graph states on the matching dimensions modulo local Cliffords  as stated by Lemma~\ref{lemma: Stab_Graph_relation}. Therefore, any stabilizer state representative on a Clifford orbit can be written as a product of representative graph states, clustering their dimensions, within their respective orbits. 

Due to Lemma~\ref{lemma: Equivalence_Graphs}, the local Clifford orbits of graph states are in one-to-one correspondence to $\{\mathrm M, \mathrm L\}$ orbits of adjacency matrices, that is, as a set, $[\mathrm{GRAPH}_\mathbf x]_\mathrm{LC} \cong [A_{|\mathbf x|}]_{\mathrm M, \mathrm L}$.

Now, note that given two vertex lists of dimensions $\mathbf d_1 = (d_1, d_1, \cdots, d_1)$ and $\mathbf d_2 = (d_2, d_2, \cdots, d_2)$ with length $n$ and adjacency matrices $A_1 \in \mathbb F_{d_1}^{n \times n}$, $A_2 \in \mathbb F_{d_2}^{n \times n}$, the following relation holds
\begin{equation}
    |G(A_1 \oplus A_2)\rangle  = |G(A_1)\rangle \otimes |G(A_2)\rangle\;.
\end{equation}
In general, $|G(\oplus_i A_i)\rangle = \otimes_i |G(A_i)\rangle$. Therefore, Eq.(\ref{eq:stabtograph}) can be written as:
\begin{equation}
G(A)\in \bigotimes_{\mathbf x \in [\mathbf d]}[\mathrm{GRAPH}_\mathbf x]_\mathrm{LC} \; \Leftrightarrow \; A \in \bigoplus_{\mathbf x \in [\mathbf d]}[A_{|\mathbf x|}]_{\mathrm  M, \mathrm L}\;.
\end{equation}
    
\end{proof}

\section{On the hardness of the optimization problem}

Since Theorem \ref{theo: main_result} explicitly formulates the stabilizer scenario as an optimization problem, we can investigate its computational hardness, which becomes particularly relevant as the number of parties increases. If the graph state is fixed, the resulting optimization problem involves only the complex variables that define the measurements/unitaries and it can always be expressed in the following form:
\begin{align}
    \max_{u \in \mathbb C^{N}} &f(u)\;, \label{eq:polyopt}\\
    & g_k(u) \geq 0 \quad ;\quad 1 \leq k \leq K
\end{align}
where $f, g_k$ are all polynomials with fixed degree, the variables $u$ corresponds to the entries of the unitaries and $K$ depends on the measurement choices scaling as $K = O(n)$. This is known as a \emph{commutative} polynomial optimization problem, or an instance of semialgebraic optimization, known to be NP-hard in the generic case \cite{laurent2009sums} since various NP-complete decision problems can be formulated as minimization/maximization of a polynomial over polynomial inequalities.

The complexity of counting all classes of graph states was found only in the case of qubits  ($\mathbf d = (2,2,\cdots,2)$) \cite{dahlberg2020clifford} . It was found that it is $\#\mathrm{P}$-complete: It is hard as counting the number of solutions of problems in $\mathrm{NP}$: we do not expect polynomial time algorithms to solve it, since otherwise $\mathrm P  =\mathrm{NP}$.

This complexity is expected, as the number of graph or stabilizer states grows exponentially with the number of parties, even when considering only locally inequivalent orbits. However, these complexity-theoretic results are only relevant in the context of scaling the number of parties. If the number of parties, along with other scenario parameters, remains fixed, the problem reduces to a standard optimization task that can be efficiently solved on a modest laptop. In the case of qubit bipartite inequalities, extensively discussed in the main text, all polynomials have a degree of at most $2$, allowing Eq. (\ref{eq:polyopt}) to be reformulated as a Quadratically Constrained Quadratic Program (QCQP), for which standardized numerical solvers are readily available.

The graph counting problem can be efficiently handled when the number of parties is small. In the bipartite case with $ \mathbf{d} = (d, d) $ for prime $d$, there are only two equivalence classes: one corresponding to a graph with an edge and the other to a graph without edges. Since $\mathrm{M}$, as defined in Eq.~(\ref{eq:Mmove}), can assign any weight to a graph with an edge, all such graphs belong to the same orbit. This aligns with the result that every stabilizer state is LC-equivalent to either $|00\rangle$ or the maximally entangled EPR pair, $ |\mathrm{EPR}\rangle = \frac{1}{\sqrt{d}} \sum_{i=0}^{d-1} |ii\rangle $.

As a slightly more complex example, consider the tripartite scenario with \( \mathbf{d} = (2,2,2) \). In this case, the adjacency matrices are binary, i.e., $ A \in \mathbb{F}_2^{3\times 3} $, and the only operation within the equivalence class $[\cdot]_\mathrm{L} $ is local complementation at each vertex $v$.  

There is a single graph with no edges, which remains invariant under local complementation. For a single edge, there are three possible graphs, corresponding to the three choices of edge placement. Since each graph contains only one edge, they are also invariant under local complementation, as $A_{iv}A_{vj} = 0$ for any choice of vertex $ v $ and probe vertices $ i $ and $j $, thereby defining three distinct LC classes.  

For two edges, there are $\binom{3}{2} = 3$ possible graphs, and for three edges, there is the complete graph $ K_3 $. Notably, these graphs are related by local complementation: from Eqs. (\ref{eq:LC1}, \ref{eq:LC2}), applying local complementation at each of  the vertices $ \{1,2,3\} $ in $ K_3 $ generates each of the two-edge graphs. Conversely, due to the binary nature of $ \mathbb{F}_2 $, performing local complementation on any vertex of a two-edge graph yields the complete graph.  Thus, we conclude that there are$ 1+3+1 = 5 $ distinct Local Clifford classes of stabilizer states for three qubits.

\section{Witnessing Magic with a three-party Bell inequality.}
In this section we show that  the inequality ~\cite{PhysRevLett.108.110501}
\begin{equation}
   B = S_3+R_2 \leq 6, \label{eq: S3_R2}
\end{equation}
 is respected by all three-qubit stabilizer states. The inequality is composed by two terms: the term $R_2$ which involves only two-body expectation values 
\begin{align}
    R_2=\sum_{i=0,1}\braket{A_1^{i}A_2^{i\oplus1}  + A_1^{i}A_3^{i\oplus1}  +A_2^{i}A_3^{i\oplus1} }, \label{eq:R2}
\end{align}
and $S_3$ which is the Svetlichny polynomial \cite{PhysRevD.35.3066}
\begin{equation}
    S_3 = \left\langle \sum_{i=0,1}(-1)^iA_1^iA_2^iA_3^i +\sum_{cyc_1}A_1^iA_2^jA_3^k -\sum_{cyc_2}A_1^iA_2^jA_3^k \right\rangle, \label{eq: svetlichny}
\end{equation}
As mentioned in the main text, the $3$-qubit pure stabilizer states are divided into three different equivalence classes of entangled states. One class can be represented by the state $\ket{000}$ (fully separable class), the second class corresponds to the bi-separable one, for which $\ket{\Phi^+}\ket{0}$ may be picked as a representative, and  the last class is the genuine entangled class, which may be presented by the state $\ket{\text{GHZ}} = \frac{1}{\sqrt{2}}(\ket{000} + \ket{111})$. 

Using the fact that all Bell inequalities are invariant under unitaries, $I(\rho) = I(U\rho U^\dagger)$, showing that these representatives respect Ineq.~(\ref{eq: S3_R2}) implies that all vertices of the stabilizer polytope will satisfy it and by an argument of convexity one can extend the proof to all stabilizer states. In principle we need to consider all the five LC classes: the above-mentioned three classes of entangled states plus the two biseparable classes obtained by considering a maximally entangled state between $A$ and $C$ and $B$ and $C$. However, since the inequality is symmetric under the permutation of parties, it is enough to consider only the first three classes of entangled states. 

To begin with, since the state $\ket{000}$ is separable, it trivially satisfies Ineq.~(\ref{eq: S3_R2}) and any other well defined Bell inequality. In turn, Ref.~\cite{PhysRevLett.108.110501} proved that this inequality is also fulfilled for any measurements on a GHZ state $\ket{\text{GHZ}}$. We are thus left to prove that the state $\ket{\Phi^+}\ket{0}$ (and its permutations) satisfies Ineq.~(\ref{eq: S3_R2}). But this is easy since the Svetlichny polynomial can achieve at most the value of $S_3=4$ with measurements on any biseparable state \cite{PhysRevD.35.3066} and, as we show next, the maximum achievable value by other term is $R_2=2$ . To see the latter, note that by considering the state $\ket{\Phi^+}\ket{0}$, all correlations $\braket{A^i_1A^k_3} =\braket{A^j_2A^k_3}=0$. This is clear since each of these terms is the average of a dichotomic operator (which can be decomposed into two projectors) on either the state $\rho_{A_1} =\text{Tr}_{A_2}(\rho_{A_1A_2})$ or the state $\rho_{A_2} =\text{Tr}_{A_1}(\rho_{A_1A_2})$, in which $\rho_{A_1A_2} = \ket{\Phi^+}\bra{\Phi^+}$. Since both reduced states are equal and proportional to the Identity operator the expectation value gives zero. The remaining terms are of the form $\braket{A^i_1A^j_2}$, but these expectation values are bounded by $1$. Then the maximum value that $R_2$ can take when considering a bipartite state $\ket{\Phi^+}\ket{0}$ is $2$. Since the value of $S_3$ is $4$ for biseparable state \cite{PhysRevD.35.3066}, it follows that $B(\ket{\Phi^+}\ket{0})\leq 6$. Due to the symmetry of the inequality, a completely analogous argument can be used to show that the two other LC bipartite classes satisfy this inequality. Therefore we have proved that all stabilizer pure states satisfy Ineq.~(\ref{eq: S3_R2}). 

\begin{figure}
    \centering
    \includegraphics[width=0.6\linewidth]{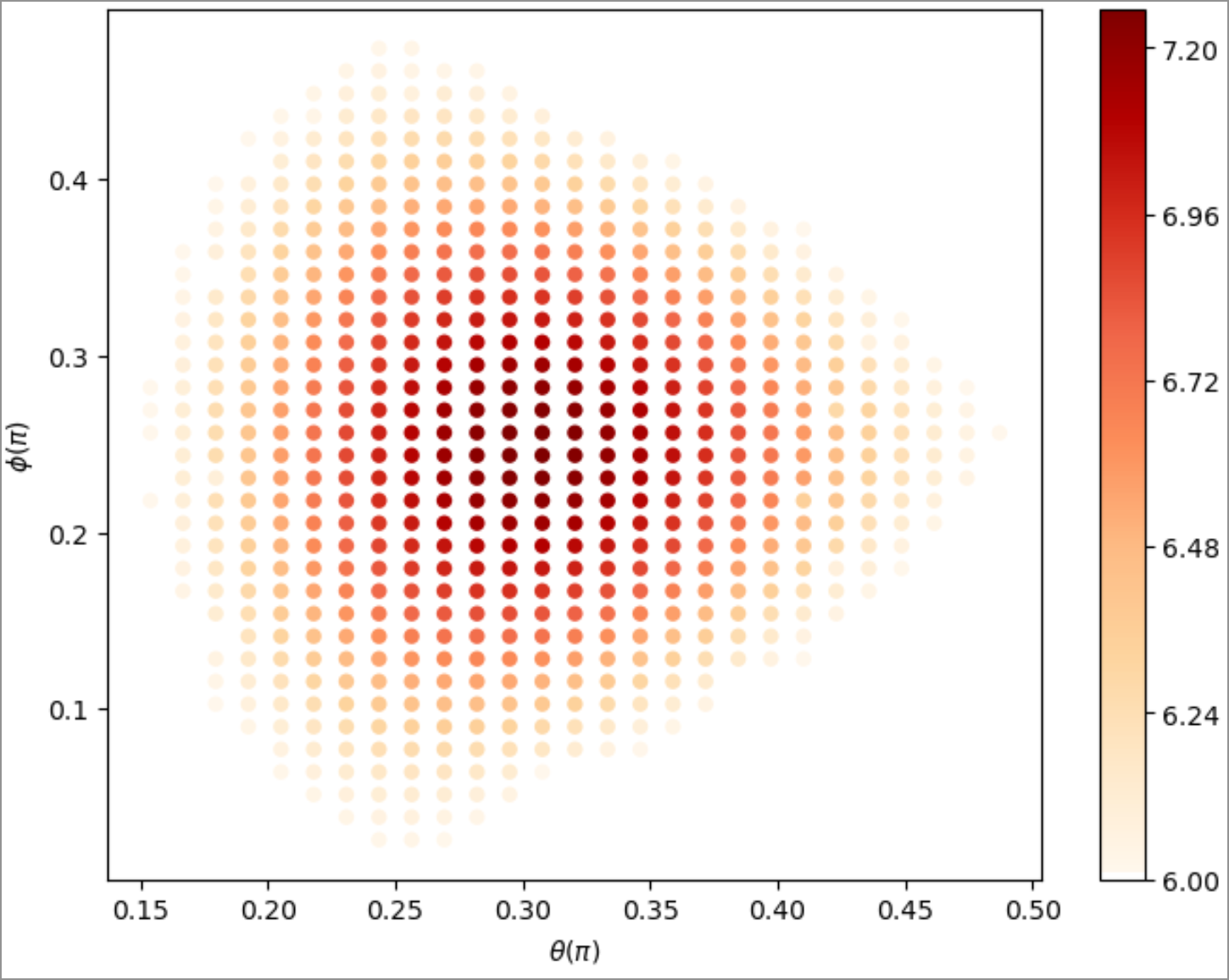}
    \caption{Violation of the Ineq.~(\ref{eq: S3_R2}) by the generalized $W$ state (\ref{eq: Gen_W}). The plot is in units of $\pi$, and the maximum is attained by the $\ket{W}$ state. }
    \label{fig: W_alpha_violations}
\end{figure}

To complement the results in the main paper, we probe the usefulness of this inequality by considering a generalized family of W-states, given by 
\begin{equation}
    \ket{W_G} = \alpha_1\ket{001}+\alpha_2\ket{010}+\alpha_3\ket{100},\label{eq: Gen_W}
\end{equation}
with the amplitude coefficients parametrized as $\vec{\alpha} = (\alpha_1,\alpha_2,\alpha_3) = (\sin(\theta)\sin(\phi),\sin(\theta)\cos(\phi),\cos(\theta))$. By performing a numerical optimization over the measurements, we search for violations of the Ineq.~(\ref{eq: S3_R2}) with the results shown in Fig.~\ref{fig: W_alpha_violations}, a heat-map as a function of $\theta$ and $\phi$. 

\bibliography{main.bib}